\documentclass[12pt,superscriptaddress]{revtex4}
\pdfoutput=1
\usepackage{amssymb,amsmath,amsthm,graphicx}
\usepackage{graphics, color}
\usepackage{subfigure} 
\usepackage{latexsym}
\usepackage{bm}
\usepackage{epsfig}
\usepackage[pdftex]{hyperref}
\usepackage{setspace}
\begin{document}
\begin{singlespacing}

\title{Rings, Ripples, and Rotation:\\Connecting Black Holes to Black Rings}
\author{\'Oscar J.~C.~Dias}
\email{oscar.dias@ist.utl.pt}
\affiliation{CAMGSD, Dept. de Matem\'atica \& LARSyS, Instituto Superior T\'ecnico, 1049-001 Lisboa, Portugal}
\affiliation{STAG research centre and Mathematical Sciences, University of Southampton, UK}
\affiliation{IPhT, CEA Saclay, CNRS URA 2306, F-91191 Gif-sur-Yvette, France}
\author{Jorge E.~Santos}
\email{jss55@stanford.edu}
\affiliation{Department of Physics, Stanford University, Stanford, CA 94305-4060, U.S.A.}
\affiliation{DAMTP, Centre for Mathematical Sciences, University of Cambridge, Wilberforce Road, Cambridge CB3 0WA, UK}
\author{Benson Way}
\email{b.way@damtp.cam.ac.uk}
\affiliation{DAMTP, Centre for Mathematical Sciences, University of Cambridge, Wilberforce Road, Cambridge CB3 0WA, UK}
\begin{abstract}\noindent{
Singly-spinning Myers-Perry black holes in $d\geq 6$ spacetime dimensions are unstable for sufficiently large angular momentum.  We numerically construct (in $d=6$ and $d=7$) two new stationary branches of lumpy (rippled) black hole solutions which bifurcate from the onset of this ultraspinning instability.  We give evidence that one of these branches connects through a topology-changing merger to black ring solutions which we also construct numerically.  The other branch approaches a solution with large curvature invariants.  We are also able to compare the $d=7$ ring solutions with results from finite-size corrections to the blackfold approach, finding excellent agreement.
}
\end{abstract}
\maketitle
\end{singlespacing}

\emph{\bf  Introduction.}
As expressed by John Wheeler's statement, ``Black holes have no hair" \cite{mtw}, black holes (BHs) in four spacetime dimensions are remarkably simple objects.  The topology, rigidity, uniqueness, and no-hair theorems ensure that Kerr BHs are the only stationary, vacuum, and asymptotically flat solutions to general relativity, and that they are uniquely specified by their mass $M$ and angular momentum $J$ \citep{Robinson:2004zz}.  Because Kerr BHs are also linear mode stable \cite{Whiting:1988vc}, any stellar object undergoing gravitational collapse towards a BH is expected to settle to the Kerr solution. 

Yet, the uniqueness of the Kerr BH may seem surprising given the close connection between gravity and fluids. According to the membrane paradigm \cite{citeulike:581004}, external observers will find that BH horizons behave like fluid membranes, endowed with a viscosity, conductivity, temperature, entropy, etc.  Moreover, in certain circumstances there are formal mappings between solutions of general relativity and solutions of Navier-Stokes \cite{Bhattacharyya:2008jc,Bredberg:2011jq}.  Fluids, however, lack strong uniqueness theorems and admit a rich structure of solutions.  Indeed, rotational instabilities appear in fluid droplets and non-spherical solutions develop \cite{Eggers:1997zz}. In particular, a ring configuration is preferred for high spin. Therefore, it may be natural to suspect that BHs behave like fluid droplets and have a greater diversity of solutions. This is the emerging picture in $d>4$ spacetime dimensions.

In higher dimensions, gravity becomes weaker as it spreads out over extra dimensions.  Horizons are therefore more flexible, creating new gravitational phenomena with no four-dimensional counterparts.  For example, in $d\geq 5$ black strings exist and suffer from the Gregory-Laflamme instability \cite{Gregory:1993vy}.  This instability leads to a fractal-like array of spherical BHs and cosmic censorship violation \cite{Kudoh:2004hs,Lehner:2010pn}.  This behaviour is similar to the Rayleigh-Plateau instability, where a fluid jet breaks into an array of spherical droplets  \cite{Cardoso:2006ks}. 

In addition, there are black rings with horizon topology $S^1\times S^{d-3}$. These have been constructed in closed analytic form in $d=5$ \cite{Emparan:2001wn} and numerically in $d=6$  \cite{Kleihaus:2012xh}. (The list of other BH horizon topologies that are allowed in $d>4$ were discussed in \cite{Cai:2001su}).  In any $d\geq 5$, black rings can be found perturbatively using the blackfold approach if the $S^1$ radius is much larger than the  $S^{d-3}$ radius \cite{Emparan:2007wm}.  In addition to new topologies, these rings introduce non-uniqueness since there are different ring solutions with the same $M$ and $J$.

Given this rich structure, understanding the space of BH solutions in higher dimensions is an open and difficult task.  Though, it is also an important task in scenarios where higher dimensions are unavoidable, such as those arising from string theory or holography.  

Let us therefore summarise of the state of the art, focusing on asymptotically flat stationary solutions.  In $d=4$ Kerr BHs are the only such solutions. For a fixed mass $M$, increasing the angular momentum $J$ decreases the BH temperature $T_H$ and the horizon area $A_H$, until the BH reaches extremality at $J=G M^2$, where the BH has vanishing temperature but non-vanishing horizon area \cite{Emparan:2008eg}.  

The higher dimensional analogues to Kerr BHs are the Myers-Perry (MP) solutions \cite{Myers:1986un}.  Also in higher dimensions, BHs can have $\left\lfloor \frac{d-1}{2}\right\rfloor$ independent angular momenta.  For simplicity, we only consider the singly-spinning case with one non-zero $J$.

Like Kerr BHs, singly-spinning MP BHs in $d=5$ at fixed $M$ will have decreasing $T_H$ and $A_H$ with increasing $J$.  Unlike Kerr BHs however, both $T_H$ and $A_H$ vanish at extremality \cite{Emparan:2008eg}, giving a naked singularity. This singularity is the merger point between the MP solutions and the $S^1\times S^2$ black rings \cite{Emparan:2001wn}.  As mentioned earlier, these rings are not uniquely specified by $M$ and $J$.  There are two branches: the ``fat" branch connected to the naked singularity, and a ``thin" branch which for large $J$ resembles bent black strings.  There is also an infinite family of solutions with disconnected horizons, such as black saturns \cite{Elvang:2007rd} and black di-rings \cite{Iguchi:2007is}, which all connect to the black ring and MP BH at the naked singularity. 

In $d\geq 6$, singly-spinning MP BHs can have arbitrarily large angular momentum. For large rotation, the horizon becomes thin as it spreads along the plane of rotation.  These BHs can become unstable to the ultraspinning instability, which is of Gregory-Laflamme type \cite{Emparan:2003sy,Dias:2009iu,Dias:2010maa}. It was conjectured that the threshold mode of this instability signals a bifurcation to a new branch of axisymmetric rotating BHs with lumpy or rippled $S^{d-2}$  horizons.  These ``lumpy" BHs are also conjectured to connect to black rings \cite{Emparan:2003sy,Emparan:2007wm}.  More lumpy BH solutions may appear from the threshold of higher harmonic modes and connect to black saturns and di-rings, etc. 

In this manuscript, we will take a firm step towards the completion of the phase diagram of singly-spinning asymptotically flat solutions of general relativity in $d\geq 6$. We confirm the existence of lumpy BHs by explicit numerical construction in $d=6$ and $d=7$. We also numerically construct black rings in $d=6$ (reproducing \cite{Kleihaus:2012xh}) and in $d=7$.  We find that there are \emph{two} families of lumpy BHs that bifurcate from the same MP solution.  Our results give robust evidence that one of these lumpy BH branches connects to the fat black ring via a topology changing merger.  The existence of the complementary branch of lumpy BHs was not anticipated in previous studies \cite{Emparan:2007wm}, and we find that they approach a solution containing large curvature.  As a byproduct of our analysis, we are also able to check the finite-size corrections to the blackfold approach for the thin black rings in $d=7$.  We find excellent agreement between these analytic results and our numerics, even for an adimensional angular momentum of order one.  

\emph{\bf  Method.}
The singly-spinning MP BHs, the lumpy BHs,  and the black rings all have an asymptotic  timelike Killing vector $\partial_t$ and rotate with angular velocity $\Omega_H$ along the rotational Killing vector $\partial_\psi$ such that $K=\partial_t+\Omega_H \partial_\psi$ generates a Killing horizon.  These geometries can be written in the form
\begin{eqnarray}\label{geometry}
& & \mathrm{d}s^2=-A\,\mathrm{d}t^2+B\,\mathrm{d}y^2+C(\mathrm{d}x+F\mathrm{d}y)^2 \nonumber \\
&& \hspace{1cm}+S_1\,(\mathrm{d}\psi-W \mathrm{d}t)^2+S_2 \,\mathrm{d}\Omega_{d-4}^2
\end{eqnarray}
where $\mathrm{d}\Omega_{d-4}^2$ is the line element of a unit $(d-4)$-sphere, and  $A,B,C,S_1,S_2,W,F$ are functions  of the coordinates $y$ and $x$. For the MP and lumpy BHs, $y$ is a radial coordinate and $x$ is an angular coordinate. For the black ring, $x$ and $y$ resemble the bipolar ring coordinates in \cite{Hong:2003gx}.  We choose these functions to vanish smoothly in the appropriate places to yield horizons with the correct topology, and demand that they have flat asymptotics.  

To find $A,B,C,S_1,S_2,W,F$, we use the DeTurck method \cite{Headrick:2009pv,Figueras:2011va}. This method requires no \emph{\`a priori} gauge fixing and yields elliptic equations for these kinds of boundary value problems \cite{Adam:2011dn}. We use pseudospectral collocation on a Chebyshev grid (a patched grid for the rings) to discretise our PDE system. The resulting system of nonlinear algebraic equations is solved using Newton-Raphson relaxation.  In our case, there are no Ricci solitons meaning the so-called DeTurck vector $\xi$ must vanish. We therefore check that $|\xi| = 0$ with an error smaller than $10^{-8}$, and that $|\xi|$ vanishes exponentially with increasing grid size, as predicted by pseudospectral methods. For the black rings, we also confirm that our results do not depend on our specific choice of gauge or patch boundary.

We compute the mass $M$ and angular momentum $J$ from a Komar Integral at infinity, and obtain the angular velocity $\Omega_H$, entropy $S_H=A_H/4$ and temperature $T_H$. We verify that the Smarr relation $\frac{d-3}{d-2}\,M=T_H\,S_H+\Omega_H J$ and the first law $\mathrm{d}M=T_H \mathrm{d} S_H +\Omega_H \mathrm{d}J$ are satisfied to less than $5\%$ error \footnote{We compute $M$ using the Smarr relation in our plots, but use the Komar integral to check our numerics.}. The lumpy BHs are parametrised by a rotation parameter in horizon radius units $a/r_+$.  The black rings are parametrised by $\Omega_H/T_H$.  More details concerning the quality of the numerics can be found in the appendix.  

\emph{\bf Results.}
The phase diagram of stationary solutions is best described by thermodynamic quantities. A meaningful comparison can be made between solutions if one introduces the dimensionless spin $j$, area $a_H$, angular velocity $\omega_H$, and  temperature $t_H$ via \cite{Emparan:2007wm}
\begin{eqnarray}\label{thermoQ}
&& j^{d-3}=c_j\, \frac{J^{d-3}}{GM^{d-2}} \,,\qquad
a_H^{d-3}=c_a\,\frac{A_H^{d-3}}{(GM)^{d-2}}\,, \\
&& \omega_H =c_\omega \, \Omega_H \, (GM)^{\frac{1}{d-3}} \,,\qquad
t_H = c_{t}\, T_H \,(GM)^{\frac{1}{d-3}}\,, \nonumber
\end{eqnarray}
where the constants  $c_j, c_a, c_\omega, c_t$ can be found in \cite{Emparan:2007wm}. 

\begin{figure}[th]
\centering
\includegraphics[width=.5\textwidth]{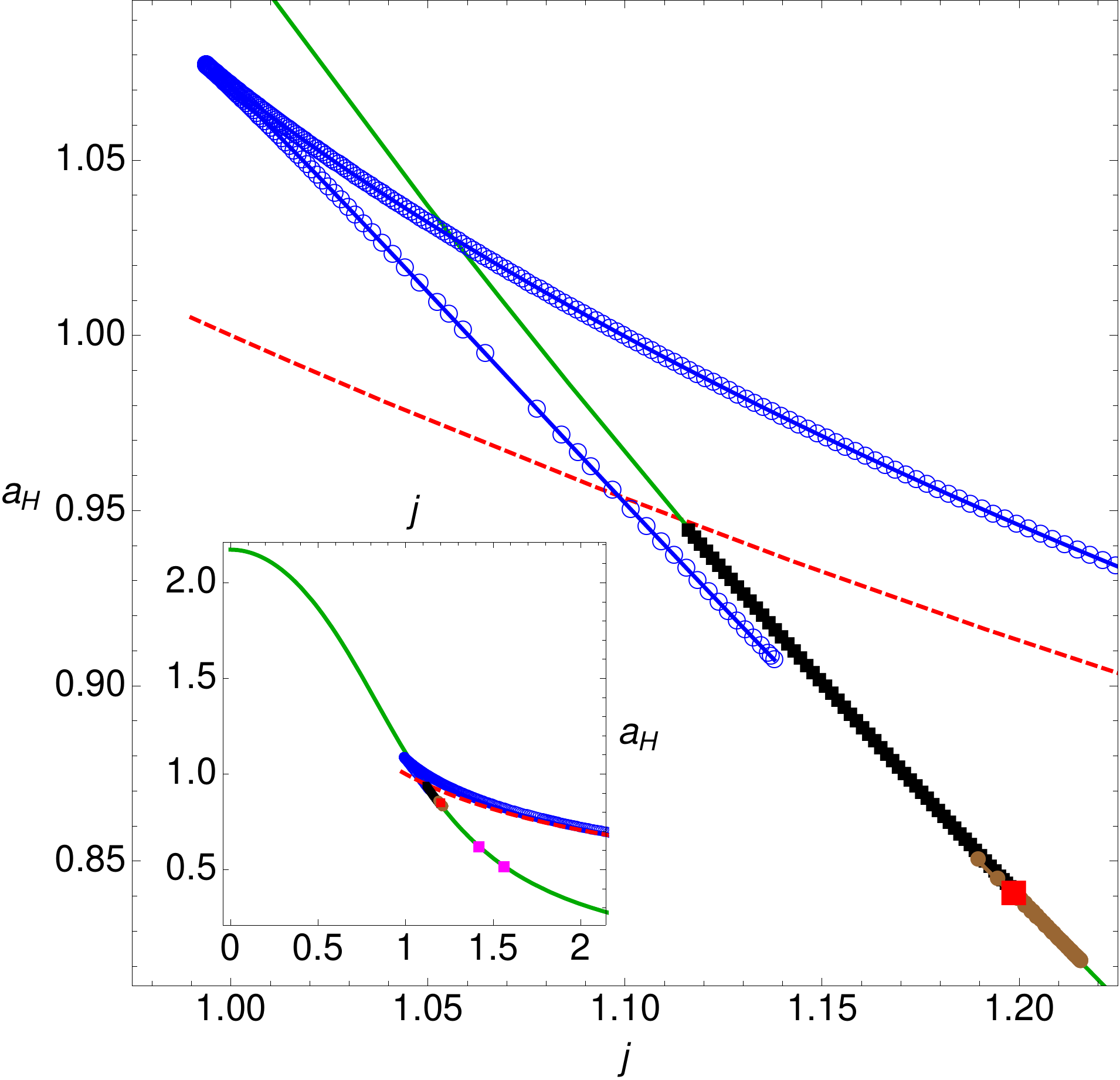}
\caption{Horizon area $a_H$ versus the spin $j$ in $d=6$. The solid green curve describes the single spinning MP BH; the large red square indicates the onset of the ultraspinning instability;  the brown dots and black squares describe the two branches of lumpy BHs.  The branch given by brown dots leads around a cusp and then towards the black rings (blue circles).  The dashed red curve is the blackfold prediction for the black rings. Isolated magenta squares on the MP BH curve are the onset of higher harmonics of the ultraspinning instability. The inset plot is a zoomed out plot.}\label{Fig:aj}
\end{figure}  
 
\begin{figure}[ht]
\centering
\includegraphics[width=.5\textwidth]{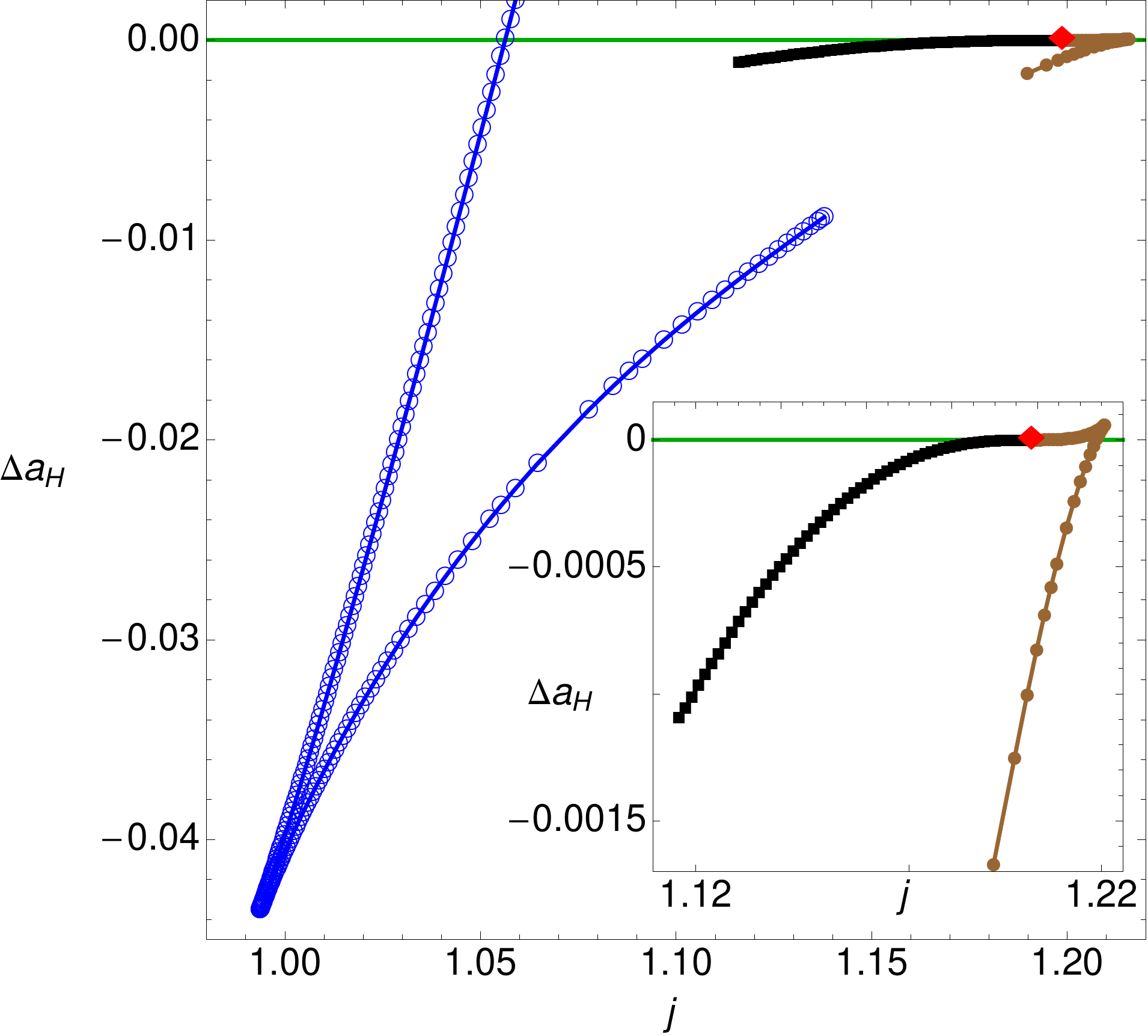}
\caption{Difference in horizon area $\Delta a_H$ between a given solution and the MP BH with the same spin $j$, as a function of $j$. (Same colour scheme as Fig.~\ref{Fig:aj}.)  The inset is a zoomed in plot.}\label{Fig:daj}
\end{figure}  

Fig.~\ref{Fig:aj} shows the horizon area $a_H$ as a function of the angular momentum $j$ in $d=6$. The solid green line describes the singly spinning MP BH \cite{Myers:1986un}. The large red square indicates the zero mode of the ultraspinning instability \cite{Dias:2009iu,Dias:2010maa}. Two families of the lumpy BHs (the black squares and brown dots) branch from this zero mode in a second-order phase transition. Note, however, that the lumpy BHs follow a curve that is close to the MP curve.  To better understand their relationship, Fig.~\ref{Fig:daj} shows $\Delta a_H$ as a function of $j$, where  $\Delta a_H$ is the difference in $a_H$ between a given solution and the MP BH with the same $j$.  

Note that the brown dots move towards a higher $j$ with a higher entropy than MP, hits a cusp, and then decreases in $a_H$ and $j$, eventually having a lower entropy than the MP BH.  This implies that the lumpy BHs are non-unique even among their own family, and provide the first example of non-uniqueness among an asymptotically flat vacuum spherical family \footnote{Such non-uniqueness can also be present in spherically charged BHs \cite{Kunz:2005ei}.}.  Continuing along this curve, we find that this branch of lumpy BHs approaches the black ring solutions given by the blue circles.  (This black ring curve extends and agrees with the curve in \cite{Kleihaus:2012xh}).  A topology-changing merger between the lumpy BHs and the black rings is expected to occur through a conical geometry \cite{Kol:2002xz,Kol:2003ja,Asnin:2006ip,Emparan:2011ve}.  Indeed, we find that the Ricci scalar of the induced horizon geometry grows as we approach the merger.

\begin{figure}[ht]
\centering
\includegraphics[width=.5\textwidth]{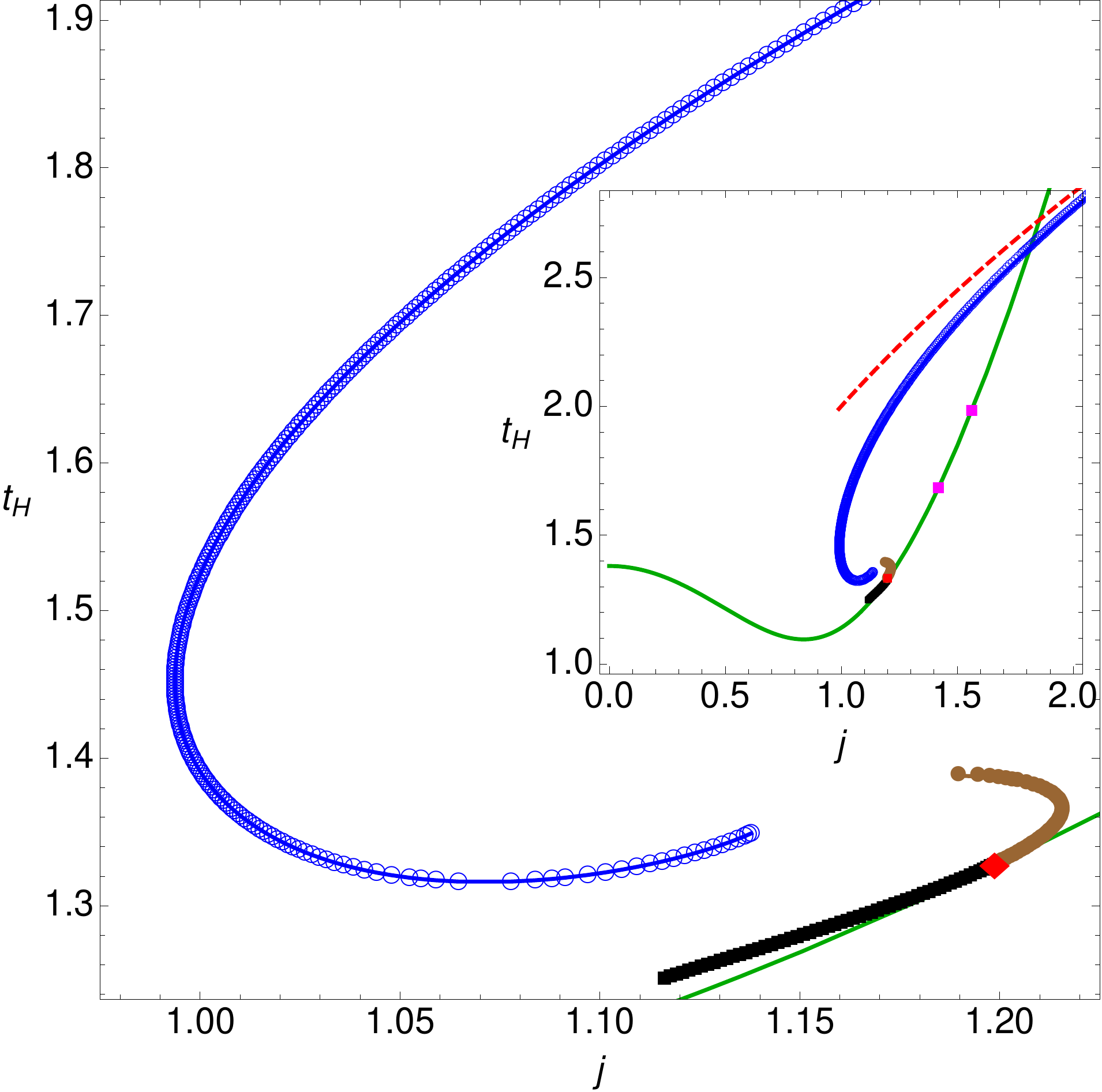}
\caption{Temperature $t_H$ as a function of the spin $j$. The inset plot is a zoomed out plot. (Same colour scheme as Fig.~\ref{Fig:aj}.)}\label{Fig:tj}
\end{figure}

\begin{figure}[ht]
\centering
\includegraphics[width=.5\textwidth]{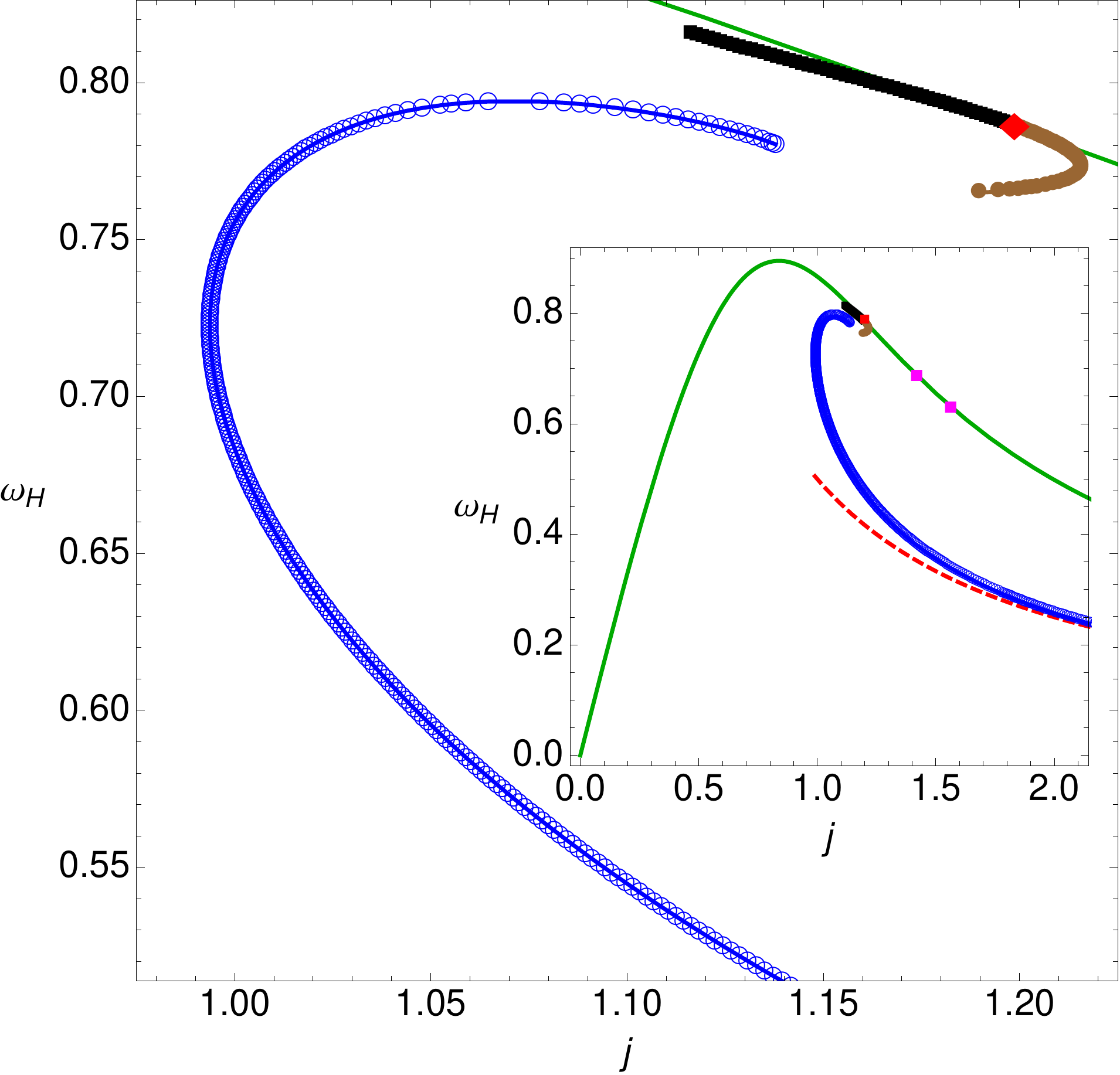}
\caption{Angular velocity $\omega_H$ as a function of the spin $j$. The inset plot is a zoomed out plot.  (Same colour scheme as Fig.~\ref{Fig:aj}).}\label{Fig:wj}
\end{figure}

As further evidence for a merger, we plot the other thermodynamic variables as a function of $j$ in Fig.~\ref{Fig:tj} and Fig.~\ref{Fig:wj}.  Again, we find the black ring and lumpy BHs approaching each other.  In these plots, the cusps are described by smooth turning points.  

The second family of lumpy BHs (the black squares) was not anticipated. To understand their existence, consider a perturbative expansion of the lumpy BHs around the ultraspinning merger point. At leading order, the amplitude of the lumpy BHs can be positive or negative and their entropy is linear in the amplitude; hence we have two branches of solutions. Note that previous studies \cite{Emparan:2007wm} drew intuition from the Kaluza-Klein (non)-uniform black string system \cite{Gregory:1993vy,Kudoh:2004hs} where periodicity ensures that the entropy is instead quadratic in the linear amplitude. 

Predicting where these lumpy BHs lead is difficult; we can only conjecture a few possibilities.  As we move along this family away from the merger with MP BH (red square), we find that curvature invariants of these lumpy BHs grow large \footnote{The Ricci scalar of the induced horizon geometry approaches $O(10^4)$ in mass units.}. The curvature is largest where the function $S_2$ in \eqref{geometry} vanishes. This suggests (but by no means implies) a possible topological transition to a $S^2\times S^{d-4}$ solution \footnote{Such solutions were found in $d=5$, but these conical singularities \cite{Figueras:2005zp}.} with rotation on the $S^2$.  Such a solution would be supported by spin-spin interaction and hence does not have a blackfold approximation at lowest order \cite{Emparan:2009vd}. We admit that there are other possibilities such as a double MP BH (also supported by spin-spin repulsion).  This lumpy BH branch might also simply end in a nakedly singular configuration.  A zero-temperature limit is also possible, but Fig.~\ref{Fig:tj} suggests we are still far from zero temperature.

\begin{figure}[ht]
\centering
\includegraphics[width=.5\textwidth]{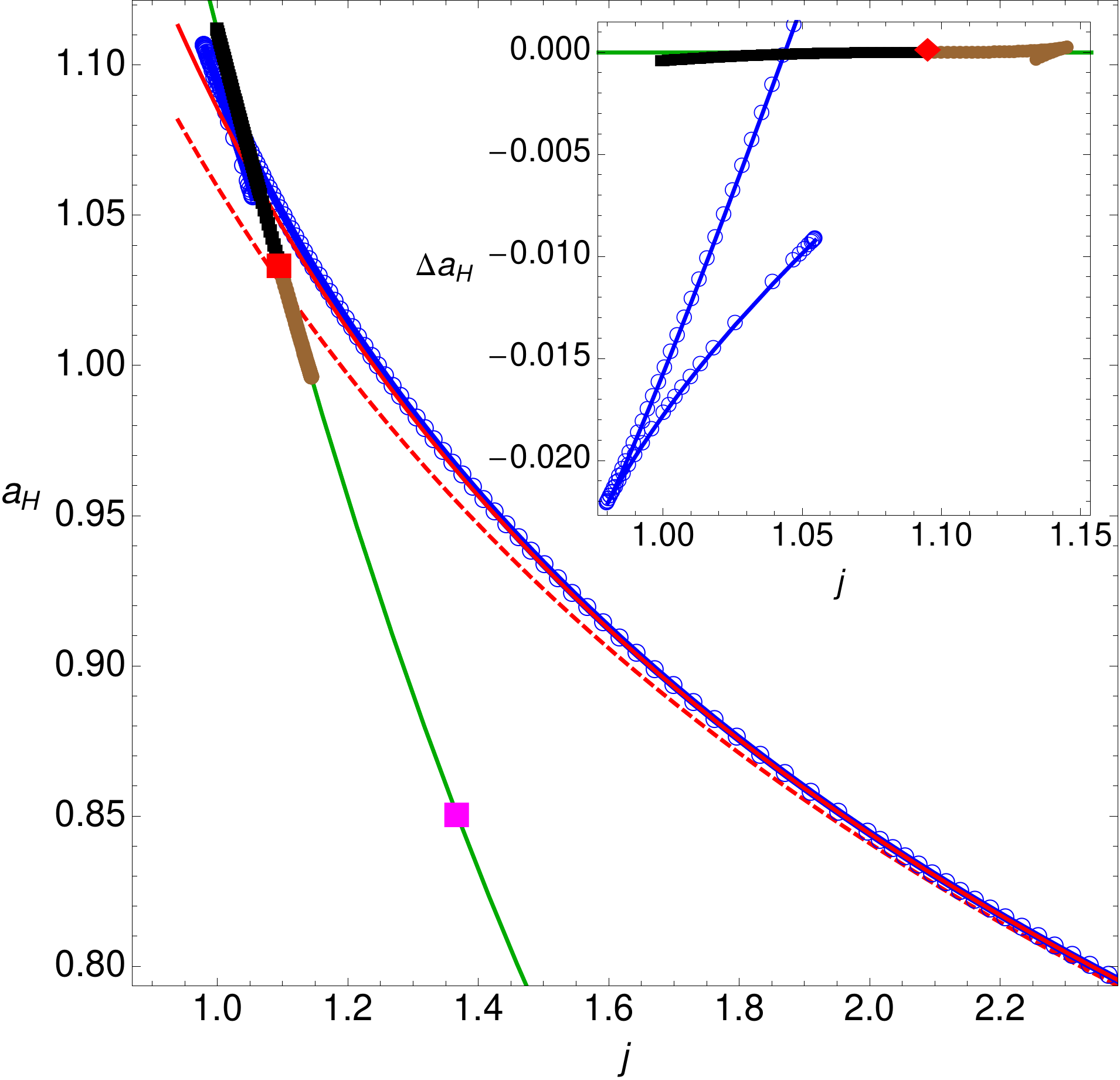}
\caption{Phase diagrams in $d=7$. The main plot is the horizon area $a_H$ as a function of the spin $j$. The inset is the difference in area $\Delta a_H$ {\it vs} the spin $j$. (Same colour scheme as  Fig.~\ref{Fig:aj}.)  The added solid red line is the blackfold curve with finite-size corrections}\label{Fig:aj7d}
\end{figure}

\begin{figure}[ht]
\centering
\includegraphics[width=.5\textwidth]{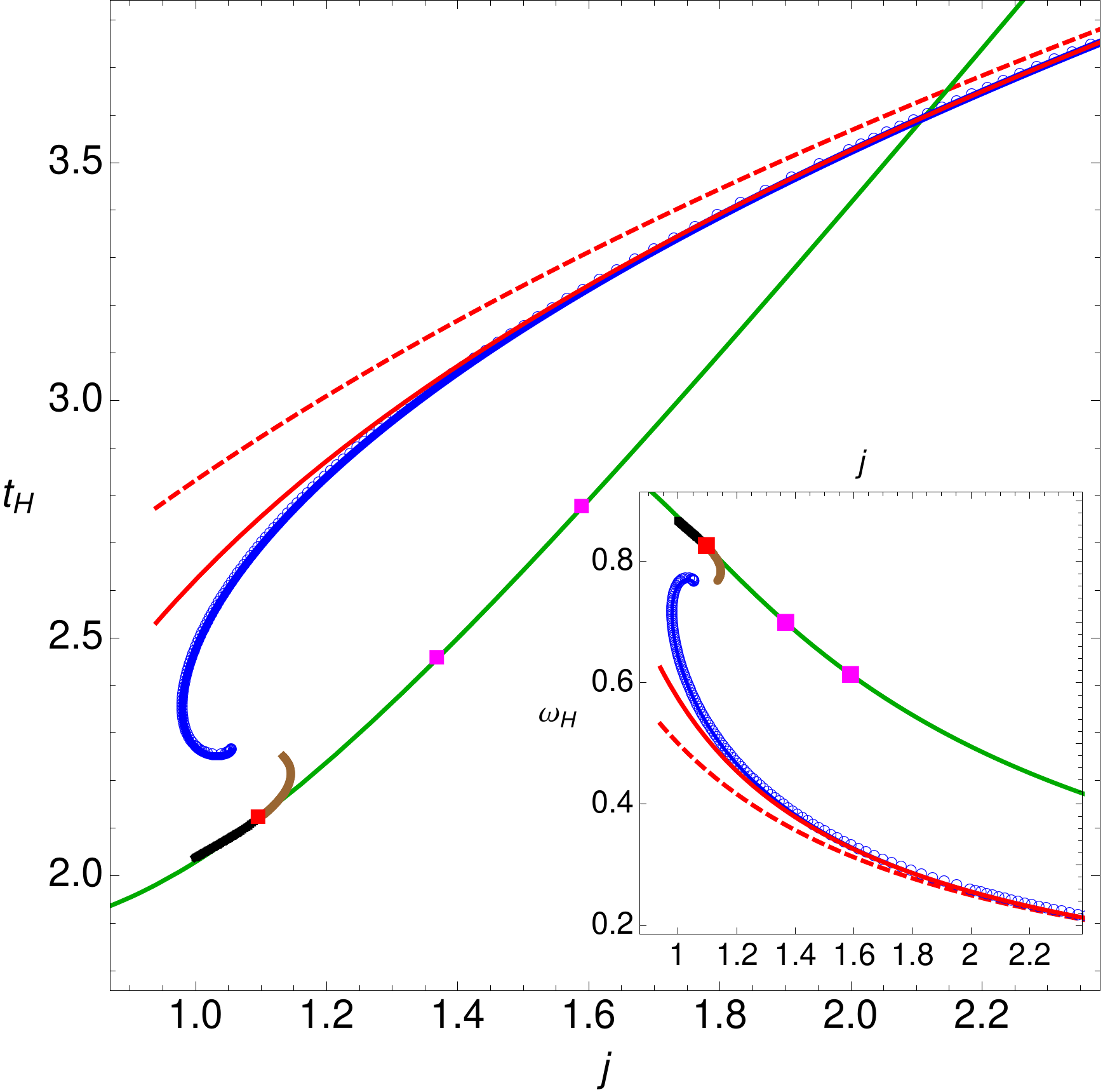}
\caption{Temperature $t_H$ as a function of spin $j$.  The inset shows the angular velocity $\omega_H$ as a function of $j$.  (Same colour scheme as  Fig.~\ref{Fig:aj7d}.)}\label{Fig:omegatj7d}
\end{figure}

We have repeated our calculation in $d=7$ and find similar behaviour.  The phase diagrams are displayed in Fig.~\ref{Fig:aj7d} and Fig.~\ref{Fig:omegatj7d}.

We can compare our numerical black ring results to analytical results from the blackfold approximation, which is valid for large $j$.  In $d=6$, this is given by $a_H\simeq \frac{1}{j},\, t_H \simeq 4 j,\,\omega_H\simeq\frac{1}{4 j^2}$, and is shown by the dashed red line in Figs \ref{Fig:aj}, \ref{Fig:tj} and \ref{Fig:wj}.  These results agree with our numerics for large $j$.  In $d>6$, one can also include finite-size corrections to the blackfold approximation (which only become dominant over self-gravitational effects in $d>6$ \cite{Armas:2011uf}).  For $d=7$, these are given by \cite{Armas14}
\begin{eqnarray}
\label{blackfold7d}
&&\hspace{-0.5cm}  \omega_H =\frac{1}{2j}\left(1+\frac{6\xi_0}{j^{8/3}}\right), \qquad t_H=\frac{3j^{1/3}}{2^{\tfrac{1}{12}}}\left(1-\frac{3\xi_0}{j^{8/3}}\right), \nonumber \\
&&\hspace{-0.5cm}  a_H =\frac{2^{\tfrac{1}{12}}}{j^{1/3}}\left(1+\frac{\xi_0}{j^{8/3}}\right),\qquad \hbox{where} \qquad \xi_0= \frac{13\Gamma(\tfrac{4}{3})^2}{(120)\big(2^{\tfrac{2}{3}}\big)\big(3^{\tfrac{1}{2}}\big)\Gamma(\tfrac{5}{6})^2}.
\end{eqnarray}
This is illustrated for $d=7$ in  Fig.~\ref{Fig:aj7d} and Fig.~\ref{Fig:omegatj7d}, where the dashed red line is the leading order blackfold result and the solid red line includes finite-size corrections.  With the corrections, the agreement is impressive, even giving $a_H$ to $\sim 1\%$ when $j\sim {\mathcal O}(1)$.  This is the first time these corrections have been compared to a numerical solution.

Finally, note that in Figs ~\ref{Fig:aj}, ~\ref{Fig:tj}, ~\ref{Fig:wj},~\ref{Fig:aj7d}, and ~\ref{Fig:omegatj7d}, the isolated magenta squares describe the zero modes of the ultraspinning higher harmonics \cite{Dias:2009iu,Dias:2010maa}. Given our results, we conjecture that two families of lumpy solutions will branch from each of these points, one of which will connect to (possibly lumpy) black saturns or black di-rings etc. 

\smallskip
\noindent{\bf Acknowledgments.}
We thank Jay Armas and Troels Harmark for kindly sharing their unpublished result \eqref{blackfold7d} \cite{Armas14}. BW thanks Joan Camps for helpful discussions and for bringing to our attention the work of \cite{Armas14}.  OJCD was supported in part by the ERC Starting Grant 240210 - String-QCD-BH.  J.E.S.'s work is partially supported by the John Templeton Foundation. BW was supported by European Research Council grant no. ERC-2011-StG 279363-HiDGR.  The authors thankfully acknowledge the computer resources, technical expertise, and assistance provided by CENTRA/IST. Some of the computations were performed at the cluster ``Baltasar-Sete-S\'ois" and supported by the DyBHo-256667 ERC Starting Grant. OJCD acknowledges the kind hospitality of the Yukawa Institute for Theoretical Physics, where part of this work has been done during the workshop ``Holographic vistas on Gravity and Strings", YITP-T-14- 1.

\appendix
\section{Numerical details}
\begin{figure}[ht]
\centering
\includegraphics[width=.45\textwidth]{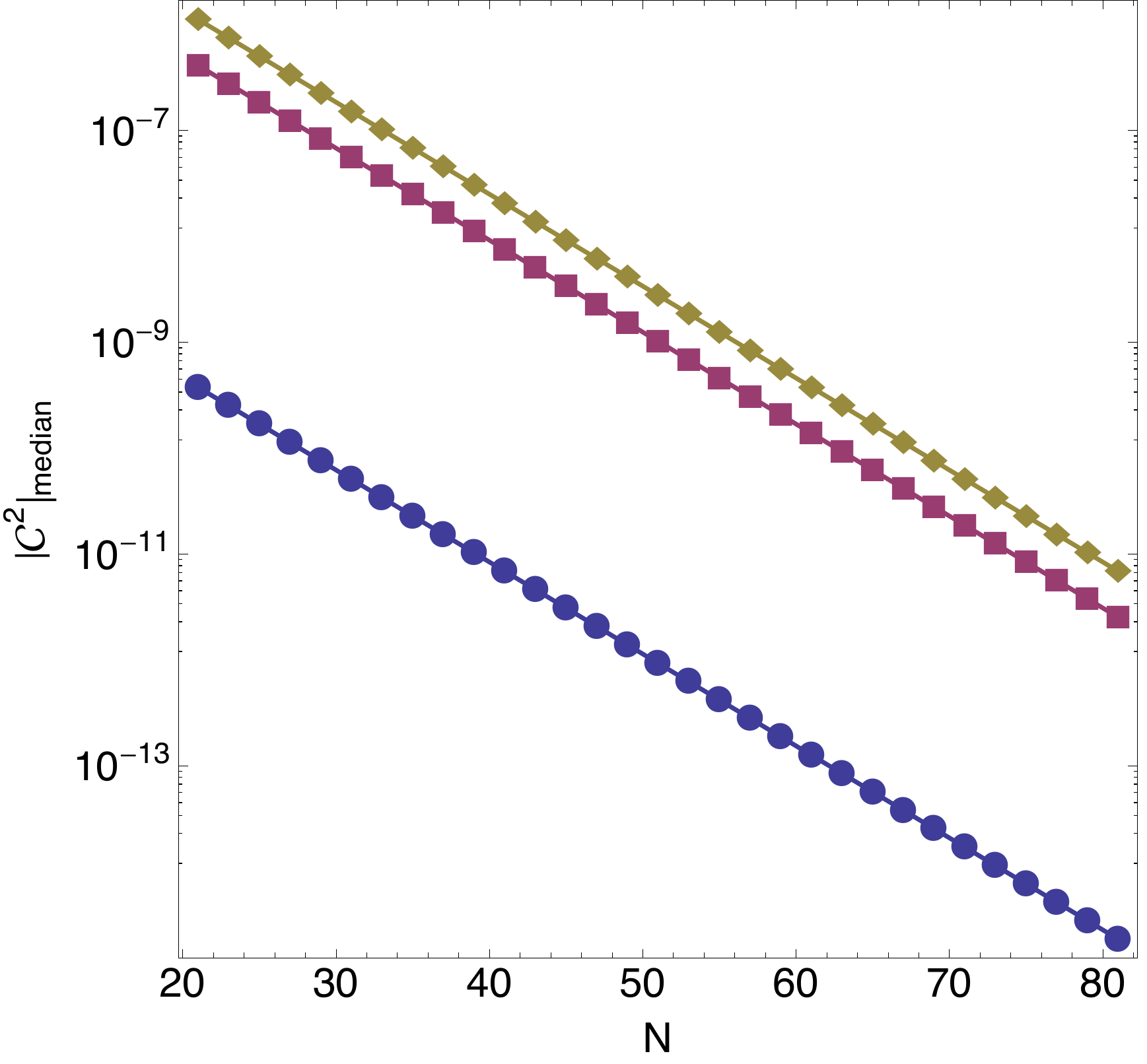}\qquad
\includegraphics[width=.45\textwidth]{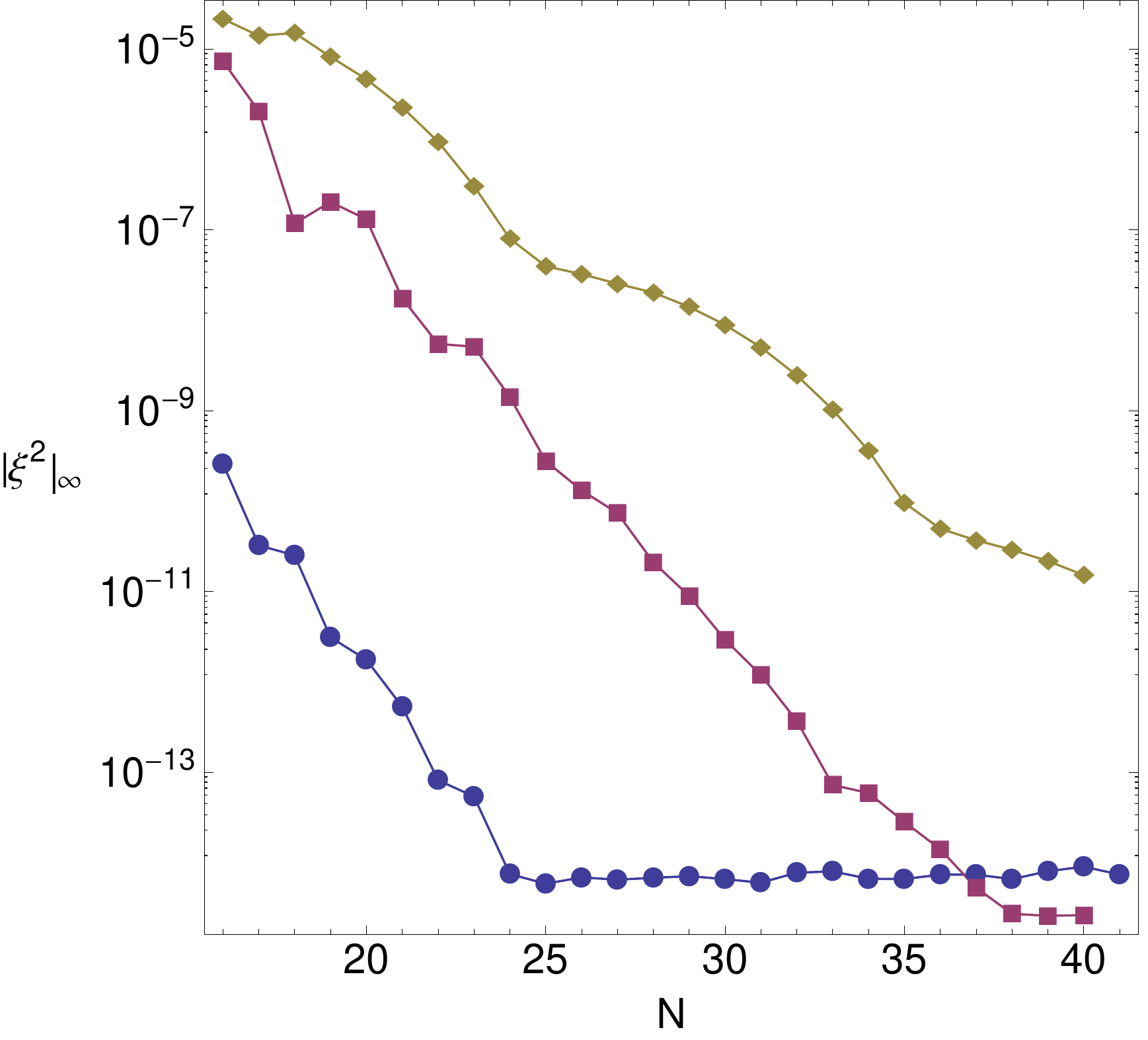}
\caption{{\bf Left:} convergence test for the Weyl tensor for lumpy black holes in $d=7$. We plot $1-|\mathcal{C}^2(N)|_\infty/|\mathcal{C}^2(N+1)|_\infty$, as a function of the number of grid points $N$. From bottom to top, we have $j = 1.097,\,1.117$ and $1.136$.  {\bf Right:} convergence test of the norm of the DeTurck vector $|\xi^2|_\infty$ for rings in $d=7$ as a function of the grid points $(N+N)\times N$.  From top to bottom, $j=1.05$ (fat branch), $j=1.01$ (thin branch), $j=1.64$ (thin branch).  The $j=1.64$ plot plateaus after reaching the limits of machine precision.}\label{fig:converge}
\label{fig:convex}
\end{figure}

In this appendix we give some numerical details regarding the construction of both the lumpy black holes and black rings.  We first test numerical convergence.  Since we are using spectral collocation methods, we expect to find exponential convergence as the number of points is varied. This is exactly what we see in Fig.~\ref{fig:converge}.  On the left panel of Fig.~\ref{fig:converge}, we consider some lumpy BHs and show how the square of the Weyl tensor $\mathcal{C}^2 = C^{abcd}C_{abcd}$ varies as the number of points is changed.

Another quantity we can use to test convergence is the norm of the deTurck vector $\xi^2$.  In the DeTurck method \cite{Headrick:2009pv,Figueras:2011va} this is the vector $\xi^\mu=g^{\alpha\beta}\left(\Gamma^{\mu}_{\alpha\beta}+\bar\Gamma^{\mu}_{\alpha\beta}\right)$, where $\bar\Gamma^{\mu}_{\alpha\beta}$ is the Levi-Civita connection for a chosen reference metric $\bar g$.  For the boundary value problems considered here, the solutions found are necessarily solutions of the vacuum Einstein equations in the gauge $\xi^\mu=0$, so the norm of the deTurck vector is a measure of how well the gauge condition is satisfied.  On the right panel of Fig.~\ref{fig:converge}, we take some black ring solutions and we plot the norm of the DeTurck vector as a function of grid points.  We again see exponential convergence.  

There is another test of numerical accuracy for the energy and angular momentum extracted from our solutions. Since our solutions are asymptotically flat, the expansion of the metric off spatial infinity is controlled by perturbations of flat space in a given gauge. We can then take advantage of the fact that flat space can be written as a warped product of a two-dimensional space (spanned by $t,R$, say) and a round $S^{d-2}$ sphere. We can thus use the expansion of \cite{Kodama:2003jz}, to catalog the possible boundary behaviours of our functions in terms of spherical harmonics on the $S^{d-2}$ sphere that preserve the $SO(2)\times SO(d-3)$ symmetry of the line element (\ref{geometry}). If we fix the metric at infinity to have the following form:
\begin{equation}
ds^2 = -dt^2+dR^2+R^2(d\theta^2+\cos^2\theta d\Omega^2_{d-4}+\sin^2\theta d\psi^2)\,,
\end{equation}
then the asymptotic behavior of $g_{tt}$ and $g_{t\psi}$ take the following simple form
\begin{align}
&g_{tt}= -1+E_0\frac{\mathbb{Y}^{0}(\theta)}{R^{d-3}}+\mathcal{O}(R^{-(d-1)})
\\
&g_{t\psi}= J_0\frac{\mathbb{Y}_{\psi}^{1}(\theta)}{R^{d-3}}+\mathcal{O}(R^{-(d-1)})\,,
\end{align}
where $\mathbb{Y}^{\ell_s}(\theta)$ and $\mathbb{Y}^{\ell_v}_a(\theta)$ are scalar and vector harmonics on $S^{d-2}$ with quantum numbers $\ell_s\geq0$ and $\ell_v\geq1$, respectively, and both $E_0$ and $J_0$ are \emph{constants}. Note that these metric functions are gauge invariant for spacetimes with the symmetries detailed above. Both the energy and angular momentum can solely be written as linear functions of these constants. 

In $d=7$, our radial coordinate is asymptotically defined as $y \propto R^{-1/2}$, which means we can extract $E_0$ by taking two numerical derivatives of $g_{tt}$.  In $d=6$, our radial coordinate is asymptotically $y\propto 1/R$, so we require three derivatives to extract $E_0$.  In any dimension, we defined our metric functions such that $J_0$ could be extracted by taking a single numerical derivative.  A good test of accuracy is to measure if these two coefficients are constants. This can be best done by performing a $\chi^2$ fit, and extracting the standard error. For all of our solutions, we find that the error in $E_0$ is smaller than $10^{-3}\%$, except for the last few points close to the merger, in which case the error increases to $0.1\%$ for the lumpy black holes, and $1\%$ for the rings.  The errors in $J_0$ are much smaller ($\sim10^{-5}\%$).

A final test can be extracted from the the Smarr law. Since we can independently compute the energy, angular momentum, angular velocity, entropy and temperature, we can test whether the Smarr law is satisfied. Again, for the solutions in $d=7$, we find that to be true within $10^{-3}\%$, except for the last couple of points close to the merger where it increases to $0.1\%$.  We find that the $d=6$ rings satisfy the Smarr law to within $10^{-3}\%$ except for the points close to the merger, where the error increases to $5\%$.  

We emphasise that the error in the Smarr law and in the energy is a generous overestimation of the error in our plots.  In our plots for the rings, the energy is obtained directly from the Smarr law after first obtaining the angular momentum, angular velocity, entropy, and temperature.  This is more accurate since it only involves taking a single derivative for the angular momentum.

\bibliography{refs}{}
\bibliographystyle{JHEP}

\end{document}